\documentclass[aps,pra,a4paper,reprint,showpacs,nofootinbib,floatfix,floats]{revtex4-1}
\usepackage{pifont}
\usepackage{color}
\usepackage{amsmath}
\usepackage{amssymb}
\usepackage{amsfonts}
\usepackage{times,txfonts}
\usepackage{bbold}
\usepackage{hyperref}
\hypersetup{
    colorlinks=true,
    citecolor=blue,
    linkcolor=blue,
    urlcolor=blue
}
\usepackage{color}
\usepackage{bm}
\usepackage{graphicx}
\usepackage{float}
\usepackage{epsfig}
\usepackage{verbatim}
\usepackage{epstopdf}
\usepackage{natbib}
\usepackage{appendix}

\providecommand{\openone}{\leavevmode\hbox{\small1\kern-3.8pt\normalsize1}}

\begin{document}
\title{
Transfer of different types of optical qubits over a lossy environment
}
\author{Hoyong Kim, Jinwoo Park, and Hyunseok Jeong}

\affiliation{Center for Macroscopic Quantum Control, Department of Physics and Astronomy, Seoul National University, Seoul 151-742, Korea}

\date{\today}

\begin{abstract}
We compare three different types of optical qubits for information transfer via quantum teleportation and direction transmission under photon losses. The three types of qubits are (1) qubits using the vacuum and the single-photon (VSP) states, (2) single-photon qubits using polarization degrees of freedom, {\it i.e.}, polarized single-photon (PSP) qubits, and (3) coherent-state qubits that use two coherent states with opposite phases as the qubit basis.
Our analysis shows that the teleportation
scheme outperforms the direct transmission for most of cases as far as fidelities are concerned.
Overall, VSP qubits are found to be the most efficient for both the direct transmission and teleportation under photon loss effects.
The coherent-state qubits are more robust than PSP qubits
either when their amplitudes are small as $|\alpha| \lesssim 1.22$ or when photon loss effects are strong.
Our results would provide useful and timely information for the development of practical optical quantum information processing particularly in the context of hybrid architectures.
\end{abstract}


\maketitle

\section{Introduction}
\label{sec:introduction}
	
Optical systems are one of the major candidates for implementations of quantum information processing.
There are different ways for qubit encoding for optical quantum information processing. Probably, the most well-known method is to use a single photon with its polarization degree of freedom. Quantum teleportation experiments have been performed using such polarized single photons (PSPs) as qubits \cite{Bouwmeester97,Boschi98} and quantum computing protocols based on linear optics have been developed along this line \cite{Knill2001,Kok07}. It is also possible to use the vacuum and single-photon (VSP) states as the basis for qubit encoding \cite{Lee00,Lund02}. Coherent-state qubits have been studied as an alternative approach to optical quantum information processing \cite{Jeong02,Ralph03} with their advantages in teleportation \cite{Enk01,Jeong01}.

Efficient transfer of qubits is an important factor in quantum information processing. It is particularly crucial for quantum communication and quantum networks \cite{Cirac97}. A comparison among the different types of qubits in terms of transfer efficiencies would be indispensable in order to build an efficient hybrid architecture for optical quantum information processing \cite{Loock,Park12,Lee13,Morin,JeongArx2013} in a lossy environment. There are different ways to transfer qubits, for example, such as direct transmission and quantum teleportation \cite{Bennett93}. Takeoka {\it et al.} compared \cite{Takeoka02} the teleportation scheme for continuous-variable states \cite{Braunstein98,Furusawa98} with the direct transmission through a noisy channel. They showed that the teleportation scheme shows better transmission performance than the direct transmission in strong decoherence regions \cite{Takeoka02}. Park and Jeong compared effects of photon losses and detection inefficiency on entangled coherent states and entangled photon-polarized states for quantum teleportation \cite{Park10}. Extending these investigations, we are interested in comparisons for both the direct transmission and teleportation with the three  aforementioned qubit-based approaches.

In this paper, we investigate and compare fidelities of information transfer for the three different types of photonic qubits over a lossy environment. We find that teleportation is more robust to photon losses than the direct transmission for VSP qubits, PSP qubits, and coherent-state qubits with small amplitudes. While VSP qubits are the most robust ones to photon losses, coherent-state qubits with small amplitudes are more robust than the PSP qubits for optical quantum information transfer.
In terms of the success probabilities for quantum teleportation based on linear optics, VSP qubits and coherent-state qubits are found to outperform PSP qubits under photon loss effects.

\section{Direct transmission and teleportation for each type of qubits}
	\begin{figure}
		\centering
		\epsfig{file=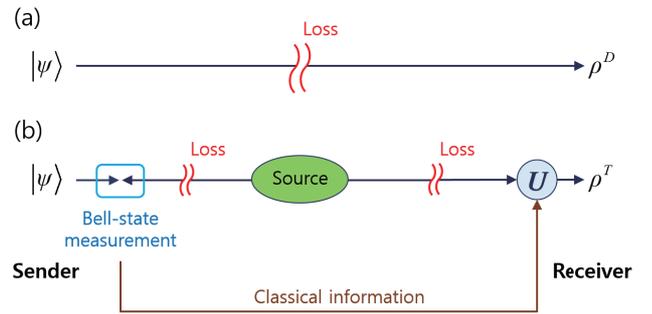,width=8.5cm}\\
		\caption{(Color online) Schematics of two different ways to transfer qubits, i.e., (a) direct transmission and (b) quantum teleportation. The state $|\psi\rangle$ represents the unknown input state, and $\rho^D$ and $\rho^T$ represent the transfered states by means of each information transfer scheme, respectively. }
		\label{scheme}
	\end{figure}

\subsection{Qubits using the vacuum and single-photon states}
	\label{sec2}
We first consider a VSP qubit,
${|\psi_{\rm V} \rangle} = \mu|0\rangle+\nu|1\rangle$,
where $|0\rangle$ and $|1\rangle$ are the vacuum and single-photon states, respectively.
This type of encoding strategy is sometimes referred to as the single-rail logic because it is defined by the occupation of a single optical mode \cite{Lee00,Lund02}.
State preperations and operations have been demonstrated experimentally using the single-rail logic \cite{Lombardi02,Resch02,Lvovsky02,Babichev04}.
The time evolution of density operator $\rho$ under photon losses is governed by the Born-Markov master equation \cite{Louisell73},
	\begin{align} \label{mastereq}
		\frac{\partial\rho}{\partial\tau}=\hat{J}\rho +\hat{L}\rho,
	\end{align}
where $\tau$ is the interaction time, $\hat{J}\rho=\gamma\Sigma_{i}a_{i}\rho a_{i}^{\dagger}$, $\hat{L}\rho=-(\gamma/2)\Sigma_{i}(a_{i}^{\dagger} a_{i}\rho+\rho a_{i}^{\dagger} a_{i})$, $\gamma$ is the decay constant, and $a_{i}$ ($a_{i}^{\dagger}$) is the annihilation (creation) operator for mode $i$. The general solution of Eq.~\eqref{mastereq} is written as, $\rho(\tau)=\exp[(\hat{J}+\hat{L})\tau]\rho(0)$, where $\rho(0)$ is the initial density operator \cite{Phoenix90}.
A VSP qubit under the direct transmission with photon losses is simply obtained as
	\begin{align}
	\begin{split}
		\rho^D_{\rm V}(\tau)=&(|\mu|^2+|\nu|^2r^2)|0\rangle\langle0|+|\nu|^2t^2|1\rangle\langle1|
		+(\mu\nu^*t|0\rangle\langle1|+{\rm H.c.}),
	\end{split}
	\end{align}
	where $\mu=\cos(\theta/2)$, $\nu=e^{i\phi}\sin(\theta/2)$, $t=e^{-\gamma\tau /2}$ and $r=\sqrt{1-e^{-\gamma\tau}}$.
	The average fidelity between input and output states is 
	\begin{align}
		F^{D}_{\rm V}(\tau)&=\frac{1}{4\pi} \int^{2\pi}_{0} \int^{\pi}_{0} \langle\psi_{\rm V} |\rho^D_{\rm V} (\tau)|\psi_{\rm V}\rangle  \sin\theta d\theta d\phi 
	=\frac{1}{2}+\frac{t}{3}+\frac{t^2}{6}.
	\end{align}

A schematic comparison between the direct transmission and the teleportation process is presented in Fig.~\ref{scheme}. In general, the quantum teleporation protocol for a qubit \cite{Bennett93} requires a bipartite entangled state as the quantum channel in addition to a Bell-state measurement scheme that discriminates the four entangled states called the Bell states. The sender's outcome for the Bell-state measurement is sent to the receiver through a classical channel so that the input state can be reconstructed by the receiver using an appropriate unitary transform ($U$ in Fig.~\ref{scheme}) \cite{Bennett93}.
 
\begin{figure}
		\centering
		\epsfig{file=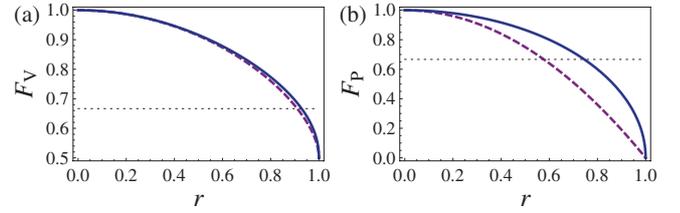,width=8.5cm}\\
		\caption{(Color online) Average fidelities of teleportation and direct transmission for 
		(a) VSP qubits (${F_{\rm V}}$) and (b) PSP qubits (${F_{\rm P}}$) against the normalized time $r$. The solid curves represent the average fidelities for teleportation  and the dashed curves correspond to those of the direct transmission. 
The horizontal dotted line indicates classical limit, $2/3$, which can be achieved by using a separable teleportation channel.}
		\label{single}
	\end{figure}

We now consider quantum teleportation of the VSP qubit using an entangled channel:
$|\Psi^-_{\rm V} \rangle=(|01\rangle-|10\rangle)/\sqrt{2}$, where $|01\rangle=|0\rangle\otimes|1\rangle$ {\it etc}.
The entangled channel at time $\tau$ is obtained using Eq.~\eqref{mastereq} as
	\begin{align}
		\rho_{\rm V}^{ch}(\tau)=t^2 |\Psi^-_{\rm V}\rangle \langle\Psi^-_{\rm V}|+r^2|00\rangle\langle00|,
	\end{align}
with which the teleportation is performed. The interaction time here should be half of the
interaction time for the direct transmission because the each part of entangled channel travels half of the length for the direct transmission as depicted in Fig.~\ref{scheme}. The Bell-state measurement is performed to discriminate between the four Bell states,
		$|\Psi ^\pm_{\rm V}\rangle=(|01\rangle\pm |10\rangle)/\sqrt{2}$ and
		$|\Phi ^\pm_{\rm V}\rangle=(|00\rangle\pm |11\rangle)/\sqrt{2}$, as a joint measurement for the input state and the sender's part of the entangled channel. 
If the outcome of the Bell-state measurement was $|\Psi^+_{\rm V}\rangle$, the required unitary transform 
is the $\sigma_z$ operation that corresponds to $\pi$ phase shift. If the outcome was  $|\Psi^-_{\rm V}\rangle$, the receiver does not need to do anything.
However, a typical Bell measurement scheme using linear optics and photodetectors \cite{Lee00} cannot discriminate the other two Bell state, $|\Phi^\pm_{\rm V}\rangle$, so that the success probability is limited to 50\% \cite{Lee00,Bjork12}.
The teleported state after an appropriate unitary transform is
\begin{align}	\label{eq:vsptel}
		\rho^T_{\rm V}(\tau)=\left(\frac{t^2}{4}+\frac{|\nu|^2 r^2}{2}\right)^{-1}\left(\frac{t^2}{4}|\psi_{\rm V}\rangle\langle\psi_{\rm V}| +\frac{|\nu|^2 r^2}{2}|0\rangle\langle0|\right),
	\end{align}
and the average success probability
is
\begin{equation}
\begin{split}
P_{\rm V}=&{\rm Tr} \left [ \langle\Psi ^+_{\rm V}| \left \{ {|\psi_{\rm V} \rangle}{\langle \psi_{\rm V} |} \otimes \rho^{ch}_{\rm V}(\tau/2) \right \}  |\Psi _{\rm V}^+\rangle \right ]_{\rm avg}\\
&+{\rm Tr} \left [ \langle\Psi ^-_{\rm V}| \left \{ {|\psi_{\rm V} \rangle}{\langle \psi_{\rm V} |} \otimes \rho^{ch}_{\rm V}(\tau/2) \right \}  |\Psi _{\rm V}^-\rangle \right ]_{\rm avg}=1/2,
\end{split}
\end{equation} 
where average is taken over the Bloch sphere of input state.
Interestingly, the success probability is not affected by photon losses even though the average fidelity is degraded as already implied in Eq.~\eqref{eq:vsptel}.
The average fidelity for the successful events obtained in the same way as the case of the direct transmission is
	\begin{align}
		F_{\rm V}^{T}(\tau/2) =\frac{1}{2(1-t)}+\frac{t^2}{4(1-t)^2}\log\frac{t}{2-t},
	\end{align}
where $\tau$  was replaced by $\tau /2$ for a comparison with the direct transmission. 	
As shown in Fig.~\ref{single}(a), the average fidelity of teleportation ${F^T_{\rm V}}$ is always higher than that of the direct transmission ${F^D_{\rm V}}$. The figure also shows that  ${F^T_{\rm V}}$ goes below the classical limit 2/3 \cite{Massar95} at $r\simeq 0.928$ while ${F^D_{\rm V}}$ does so at $r\simeq 0.910$.
	
\subsection{Polarized single-photon qubits}
	\label{sec3}
	A PSP qubit is represented as
		$|\psi_{\rm P} \rangle=\mu|H\rangle+\nu|V\rangle$,
where $|H\rangle$ and $|V\rangle$ correspond to horizontally and vertically polarized states, respectively. 
Using Eq.~\eqref{mastereq}, it is straightforward to find  that 
a PSP qubit in the direct transmission under photon losses evolves as
	\begin{align}
		\rho_{\rm P}(\tau)=t^2|\psi_{\rm P}\rangle\langle\psi_{\rm P}|+r^2|0\rangle\langle0|
		\label{eq:DP}
	\end{align}
	and the average fidelity is obtained as $F^{D}_{\rm P}(\tau)=t^2$.
We then consider quantum teleportation for a PSP qubit using an entangled channel: $|\Psi^-_{\rm P}\rangle=(|HV\rangle-|VH\rangle)/\sqrt{2}$. The entangled channel at time $\tau$ obtained using Eq.~\eqref{mastereq} is
	\begin{align}
		\rho_{\rm P}^{ch}(\tau)=t^4|\Psi^-_{\rm P}\rangle\langle\Psi^-_{\rm P}|+2r^2 t^2\tilde{\rho} +r^4|00\rangle\langle00|,
	\end{align}
	where $\tilde{\rho}=\left(|H0\rangle\langle H0|+|V0\rangle\langle V0|+|0H\rangle\langle 0H|+|0V\rangle\langle 0V|\right)/4$.
Here, the four Bell states are
		$|\Psi_{\rm P}^\pm\rangle=(|HV\rangle\pm |VH\rangle)/\sqrt{2}$ and
		$|\Phi_{\rm P}^\pm\rangle=(|HH\rangle\pm |VV\rangle)/\sqrt{2}$.
Here, the Bell-state measurement can be performed using 
a 50:50 beam splitter, two polarizing beam splitters and four photodetectors \cite{Lutkenhaus99}.
The teleported state after the Bell-state measurement and a correct unitary transform is
found to be identical to $\rho_{\rm P}(\tau)$ for the case of the direct transmission in Eq.~\eqref{eq:DP}. 
Again, only two of the Bell states, $|\Psi_{\rm P}^\pm\rangle$, can be identified using linear optics \cite{Lutkenhaus99,Calsamiglia01}
and required unitary transforms are the identity operation and the $\sigma_z$ operation that is realized with a half-wave plate.
The average success probability is 
\begin{equation}
\begin{split}
P_{\rm P}=&{\rm Tr} \left [ \langle\Psi ^+_{\rm P}| \left \{ {|\psi_{\rm P} \rangle}{\langle \psi_{\rm P} |} \otimes \rho^{ch}_{\rm P}(\tau/2) \right \}  |\Psi _{\rm P}^+\rangle \right ]_{\rm avg}\\
&+{\rm Tr} \left [ \langle\Psi ^-_{\rm P}| \left \{ {|\psi_{\rm P} \rangle}{\langle \psi_{\rm P} |} \otimes \rho^{ch}_{\rm P}(\tau/2) \right \}  |\Psi _{\rm P}^-\rangle \right ]_{\rm avg}=t/2.
\end{split}
\end{equation}
In fact, the success probability in this case is identical for any input state,
it is worth noting that the success probability of PSP qubits is always lower than the success probability of VSP qubits.
We obtain the average fidelity for the successful events as
$F_{\rm P}^T(\tau/2)=t$,
where $\tau/2$ replaces $\tau$ for a comparison with the direct transmission as mentioned in Sec.~\ref{sec2}.
We plot $F_{\rm P}^D(\tau)=t^2$ and $F_{\rm P}^T(\tau/2)=t$ in Fig.~\ref{single}(b). Obviously,
$F_{\rm P}^T$ is always higher than  $F_{\rm P}^D$, and
$F_{\rm P}^T$ goes below the classical limit 2/3 at $r\simeq 0.745$ while $F_{\rm P}^D$ does so at $r\simeq 0.577$.

\subsection{Coherent-state qubits}
	\begin{figure}[!]
		\centering
		\epsfig{file=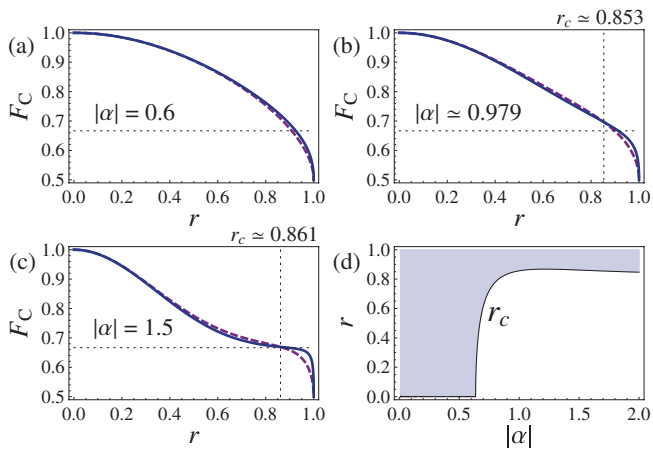,width=8.5cm}
		\caption{(Color online) (a-c) Average fidelities $F_{\rm C}$ of teleportation (solid curve) and direct transmission (dashed curve) for coherent-state qubits with amplitudes
		(a) $|\alpha|=0.6$, (b) $|\alpha| \simeq 0.979$ and (c) $|\alpha|=1.5$ against the normalized time $r$.
		The horizontal dotted lines indicate the classical limit, 2/3. (d) The shaded area indicates the region where the teleportation outperforms the direct transmission. The time boundary between the teleportation-efficient and direct-transmission-efficient regions is indicated by $r_c$.}
		\label{diffcoh}
	\end{figure}
	
Instead of single photons, superpositions of coherent states can be used for quantum information processing with their inherent advantages. The small-size implementations of superpositions of coherent states have been performed \cite{cat1,cat2,cat3,cat4,cat5,cat6} up to $\alpha \sim 1.6$ \cite{cat3,cat4,cat6} and
arbitrary qubits were demonstrated \cite{c-qubit-exp}. Their large-size implementation is possible using the non-deterministic amplification scheme \cite{Lund04}, a Fock state with a large number \cite{cat3,Lance06}, or multiple photon subtractions \cite{Dakna97,Marek09} but it is yet experimentally challenging.

Coherent-state qubits not only lose their coherence but also undergo amplitude damping under photon losses. However, as the interaction time $\tau$ is the value known to the sender and the receiver, we can use $|\pm t\alpha\rangle$ as a dynamic qubit basis in order to reflect the amplitude damping as suggested in Ref.~\cite{Jeong01}. Adopting damped coherent states $|\pm t\alpha\rangle$ as the dynamic qubit basis, the time dependent target coherent-state qubit which we want the receiver to have is
	\begin{align}	
	\label{initial_coh}
		|\psi_{\rm C}(\tau) \rangle=N(\tau)\left(\mu| t\alpha \rangle+\nu|-t\alpha \rangle\right),
	\end{align}
	where $\mu$, $\nu$ are some complex numbers and $N(\tau)$ is normalization constant. To achieve this purpose, the sender actually transmit the state $|\psi_{\rm C} (\tau=0)\rangle$. It is straightforward to find that the initial state $|\psi_{\rm C} (\tau=0)\rangle$ under direct transmission evolves to
\begin{align}
\nonumber
		\rho_{\rm C}^D(\tau)=&|N(\tau=0)|^2\Big\{ |\mu|^2|t\alpha \rangle\langle t\alpha |+|\nu|^2|-t\alpha \rangle\langle -t\alpha |\\
		&+e^{-2|\alpha |^2r^2}(\mu\nu^*|t\alpha \rangle\langle -t\alpha |+ {\rm H.c.})\Big\}.
	\end{align}
	Since the coherent states $|\pm t\alpha\rangle$ are not orthogonal to each other, we need an orthonormal basis which spans the input and the output states in order to obtain average fidelity on the Bloch sphere. We take such a basis, $|\pm(t)\rangle\propto |t\alpha \rangle\pm |-t\alpha \rangle$, 
where the normalization factors are omitted. The input state is then represented as
		$|\psi_{\rm C} (\tau) \rangle=
		\cos (\theta/2)|+(t)\rangle+\sin (\theta/2)e^{i\phi}|-(t)\rangle$
so that the average can be taken over $\theta$ and $\phi$.
The average fidelity between $|\psi_{\rm C} (\tau) \rangle$ and $\rho_{\rm C}^D (\tau)$ is obtained as
	\begin{align}\nonumber
		F^D_{\rm C}(\tau) 
		=& \frac{1}{{6\left( {{e^{4{|\alpha|^2}}} - 1} \right)}} \Big\{  - 3(1 - {e^{4{|\alpha|^2}}}) - {e^{2{|\alpha|^2}{r^2}}}(1 - {e^{4{|\alpha|^2}{t^2}}})  \\ 
		&+ \left( {{e^{4{|\alpha|^2}}} + {e^{2{|\alpha|^2}(2 - {r^2})}}} \right)\sqrt {1 - {e^{ - 4{|\alpha|^2}}}} \sqrt {1 - {e^{-4{|\alpha|^2} t^2}}} \Big\}.
	\end{align}
	
The average fidelity of teleportation was derived in Ref.~\cite{Park10} using the methods described in the previous subsections.  To perform teleportation for a coherent-state qubit, an entangled coherent state $\propto|\alpha\rangle|-\alpha\rangle-|-\alpha\rangle|\alpha\rangle$ is  shared by the sender and the receiver. The Bell-state measurement are supposed to discriminate between the states $|\Psi^\pm_{\rm C}(\tau)\rangle=N^\pm_\alpha(\tau)(|t\alpha\rangle|-t\alpha\rangle-|-t\alpha\rangle|t\alpha\rangle)$ and $|\Phi^\pm_{\rm C}(\tau)\rangle=N^\pm_\alpha(\tau)(|t\alpha\rangle|t\alpha\rangle-|-t\alpha\rangle|-t\alpha\rangle)$ where $N^\pm_\alpha(\tau)=1/\sqrt{2\pm 2e^{-4t^2|\alpha|^2}}$.
This type of Bell-state measurement can be performed using a 50:50 beam splitter and two photon-number-resolving detectors \cite{Jeong01}. The two measurement outcomes,  $|\Psi^-_{\rm C}(\tau)\rangle$ and $|\Phi^-_{\rm C}(\tau)\rangle$, require straightforward unitary transforms (identity and $\pi$ phase shift) and we take them as the successful events following Ref.~\cite{Park10}.

By substituting $\tau$ in Ref.~\cite{Park10} with $\tau/2$ for a comparison with direct transmission as mentioned in Secs.~\ref{sec2} and \ref{sec3}, the average fidelity of teleportation for the successful events is 
	\begin{align} \nonumber
	F^T_{\rm C}(\tau/2)=&\frac{1}{2}{\rm{csch }}~A~\Big\{ {\rm{csch }}~A~{{\sinh }^2}\left( {2{|\alpha| ^2} t } \right)\cosh \left( {A - 2{|\alpha| ^2}} \right)\times \\
	&{{\tanh }^{ - 1}}\left( {{\rm{csch }}{2|\alpha| ^2}\sinh A} \right) - \sinh {2|\alpha| ^2}\cosh \left( {2{|\alpha|^2} t} \right) \Big\},
	\end{align}
	where $A=2 |\alpha| ^2 \left(t-1\right)$.
The average success probability was shown to be $P_C=1/2$ \cite{Park10}. This is identical to that of VSP qubits, which is always higher than that of PSP qubits.
We plot  $F^D_{\rm C}(\tau)$ and $F^T_{\rm C}(\tau/2)$ for several amplitudes of $|\alpha|$'s in Figs.~\ref{diffcoh}(a)-(c). If the amplitudes of coherent-state qubits are small as $|\alpha|\lesssim0.636$, $F^T_{\rm C}$ is always higher than $F^D_{\rm C}$. However, as $|\alpha|$ gets larger, the region where teleportation outperforms diminishes. The direct transmission outperforms for the weaker decoherence $r<r_c$, whereas the teleportation is better for the stronger decoherence $r>r_c$ (Fig.~\ref{diffcoh}(d)).

\section{comparing different types of qubits}
	\begin{figure}
		\centering
		\epsfig{file=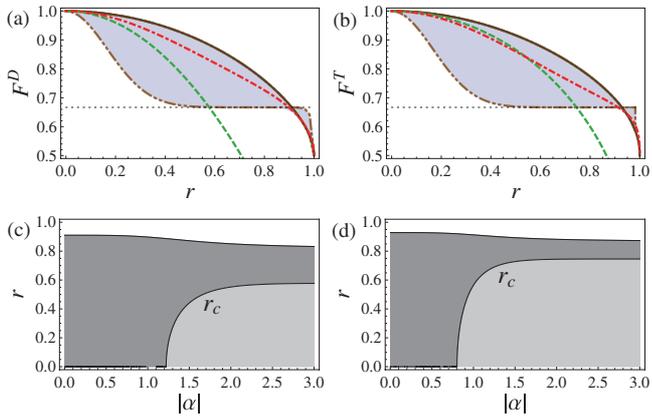,width=8.5cm}
		\caption{(Color online) The upper figures show the average fidelities for (a) direct transmission and (b) quantum teleportation
		against the normalized time $r$. The solid and  dashed curves represent the VSP and PSP qubits, respectively. 
		The dot-dashed curve corresponds to the coherent-state qubits with $|\alpha| \simeq 0.979$, and the double-dot-dashed curve to the coherent-state qubits with $|\alpha|=3$.
		The shaded area is for the coherent-state qubits with $0 \le |\alpha| \le 3$. 
The lower figures compare PSP and coherent-states qubits. The coherent-state qubits outperform PSP qubits in the dark-shaded regions while PSP qubits work better in the light-shaded regions for (c) direct transmission and (d) teleportation. In the unshaded regions of panels (c) and (d), both the fidelities are smaller than the classical bound $2/3$.}
		\label{dirtel}
	\end{figure}

	We now compare VSP qubits, PSP qubits, and coherent-state qubits under each information transfer scheme. The average fidelities for direct transmission and teleportation of VSP, PSP, and coherent-state qubits with $0 \le |\alpha| \le 3$ are plotted in Figs.~\ref{dirtel}(a) and \ref{dirtel}(b). In both the schemes, VSP qubits are the most robust ones to decoherence in the region where comparing fidelities is meaningful, {\it i.e.}, above the classical bound $2/3$. Using direct transmission (teleportation), the coherent-state qubits with small  $|\alpha| \lesssim 1.222$ ($|\alpha| \lesssim 0.802$ for the teleportation case) outperforms PSP qubits in the entire region of $r$ where the comparison is valid. However, as $|\alpha|$ gets larger, the regions where coherent-state qubits outperforms PSP qubits diminish. Coherent-state qubits outperform PSP qubits for the stronger decoherence $r>r_c$, whereas PSP qubits outperform coherent-state qubits for the weaker decoherence $r<r_c$ (Figs.~\ref{dirtel}(c) and \ref{dirtel}(d)).

Considering the number of photons as a resource,
we may compare PSP qubits and coherent-state qubits when they have the same average photon number, {\it i.e.}, $\langle \hat{n}\rangle_{\rm avg}=1$.
The average photon number of input coherent-state qubits $|\psi_{\rm C} (\tau=0)\rangle$ is
	\begin{align}
		\langle \hat{n}\rangle_{\rm avg}
				=\frac{1}{4\pi}\int^{2\pi}_{0} \int^{\pi}_{0}   \langle\psi_{\rm C}(0)|\hat{n}|\psi _{\rm C}(0)\rangle  \sin\theta   d\theta d\phi=\frac{|\alpha |^2}{\tanh (2|\alpha |^2)},
	\end{align}
	where $\hat{n}=a^{\dagger}a$.
	Therefore, the amplitude of coherent-state qubits for a comparison should be $|\alpha|\simeq 0.979$ for $\langle \hat{n}\rangle_{\rm avg}=1$ to be the same to that of the PSP qubits. 
The coherent-state qubits with the chosen amplitude $|\alpha| \simeq 0.979$ always outperform PSP qubits when the direct transmission is used as shown in Fig.~\ref{dirtel}(a). However, when the teleportation protocol is used, PSP qubits are more robust than coherent-state qubits  with the chosen amplitude when decoherence is weak, and the opposite is true for strong decoherence (Fig.~\ref{dirtel}(b)). The PSP qubits in both the schemes eventually become the vacuum states, which leads their fidelities in Fig.~\ref{dirtel} to vanish as $r \rightarrow 1$.

The coherent-state qubits and the VSP qubits become identical in the limit of $\alpha\rightarrow0$ as implied in Figs.~\ref{dirtel}(a) and \ref{dirtel}(b). This is due to the fact that even and odd superpositions of coherent sates, $|\alpha\rangle\pm|-\alpha\rangle$ (without normalization),
approach the vacuum and single photon, respectively \cite{Jeong03}.

\section{remarks}
	
Several different types of qubits have been suggested for optical quantum information processing and each of them
has its own merits and limitations.
A hybrid architecture using different types of qubits may be
an efficient way to implement practical quantum information processing based on optical systems
\cite{Loock,Lee13,Park12}.
In this context, it is important to make a thorough comparison among the different types of qubits in terms of transfer efficiencies in a lossy environment. We have compared three well-known  different types of optical qubits, VSP, PSP and coherent-state qubits, for information transfer via quantum teleportation and direction transmission under photon losses.

Of course, it should be noted that quantum teleportation always suffers lower success probabilities
compared to the direct transmission
if available resources are limited to linear optics elements and photon detectors in addition to the entangled pair \cite{Lutkenhaus99,Calsamiglia01}.
However, as far as fidelities are concerned, quantum teleportation always outperforms the direct transmission when VSP and PSP qubits are used. The same applies to the coherent-state qubits when their amplitudes are as small as $|\alpha|\lesssim0.636$. On the other hand, the teleportation outperforms the direct transmission in the strong decoherence regions for the coherent-state qubits with large amplitudes.

We have found that VSP qubits are the most robust ones against photon losses both for quantum teleportation and for the direct transmission.
Coherent-state qubits with small amplitudes ($|\alpha| \lesssim 1.222$ for direct transmission and $|\alpha| \lesssim 0.802$ for teleportation) are more robust to photon losses than PSP qubits in optical quantum information transfer, whereas the coherent-state qubits with large amplitudes outperforms PSP qubits only in the strong decoherence regions.
This means that coherent-state qubits may be more effective than PSP qubits
for optical quantum information transfer particularly when photon loss effects are heavy. The success probabilities for teleportation of coherent-state qubits and VSP qubits ({\it i.e.}, 1/2 regardless of losses) are always greater than that of PSP qubits ({\it i.e.}, smaller than 1/2 under lossy effects). Overall, VSP qubits are the most efficient for quantum information transfer under photon loss effects among the three types of qubits.

In spite of our results clearly unfavorable to the PSP qubits, the PSP qubits may be preferred for certain applications such as quantum key distribution using single photons in which post-selection plays an important role \cite{Scarani09}. In this type of post-selection process, a result is simply discarded whenever any photon is missing at the final measurement. This is not so straightforward with the VSP or coherent-state qubits because the photon numbers of those qubits are inherently indefinite.

In this paper, we have compared three types of optical qubits that can be represented by single-mode states.
Our results would provide useful and timely information for the development of practical optical quantum information processing.
It would be an interesting future work to extend this comparison to optical qudits \cite{o-qudit,c-qudit}, continuous variable systems \cite{LoydBraun}, and  hybrid qubits \cite{Lee13,Morin,JeongArx2013}.

\section*{Acknowledgements}

This work was supported by the National Research Foundation of Korea (NRF) grant funded by the Korea government (MSIP) (No. 2010-0018295) and by the Center for Theoretical Physics at Seoul National University.

\end{document}